# GROUP FORAGING IN DYNAMIC ENVIRONMENTS

Michael E. Roberts, Sam Cheesman, Patrick McMullen

DePauw University
7 E. Larabee St.
Greencastle, IN, 46135, United States
e-mail: michaelroberts@depauw.edu

## ABSTRACT

Previous human foraging experiments have shown that human groups routinely undermatch environmental resources much like other animal species. In this experiment, we test whether humans also selectively rely on others as information sources when the environmental state is uncertain, and we also test whether overt signals of other foragers' success influences group matching behavior and group adaptation to a changing environment. The results show evidence of reliance on social information in specific conditions, but participants were primarily influenced by their individual assessments of food location rather than the success of other foragers.

## INTRODUCTION

Groups of foraging animals often display undermatching in which there are proportionally fewer than expected foragers at the more abundant resource. For instance, a pool with 70% of the resources may only attract 65% of the foragers, even after normalizing based on the foragers outside of the pools. These results deviate from the ideal free distribution theory in biology (Fretwell and Lucas, 1973), which expects optimal distributions of foragers to food, i.e. 70% of the foragers should be in the 70% pool.

Recent human group foraging experiments by Goldstone and Ashpole (2004) and Goldstone, Ashpole, and Roberts (2005) used Java networked computers to allow participants to move around in a virtual world and competitively collect food in two resource pools. The experiments varied the visibility/invisibility of food and the visibility/invisibility of other foragers. In general, human groups also undermatched to the resource pools; however, a visible food/invisible foragers condition led to overmatching, in which more than the expected proportion of foragers were in an 80% pool.

Roberts and Goldstone (2006) subsequently designed an agent-based model, EPICURE, to account for the aforementioned results and explain the factors that influenced the human foragers. EPICURE accurately fit the human data, and it offered a novel explanation of undermatching behavior. Undermatching occurs when foragers' behavior distorts the information available in the environment and therefore influences subsequent foraging choices. For instance, in the previous foraging experiments, the resource pools were Gaussian distributed (with high densities of food in the center of a pool and less food on the periphery), so a forager who patrolled the center of a pool would thereby collect a disproportionate amount of food and make it difficult for others to estimate the relative proportions of food in the two pools.

EPICURE also led to new analyses of the original foraging data, and the analyses indicated that humans generally chose their movements based on their own private information (i.e., where they had recently acquired food) rather than using public information, such as the locations of other foragers. However, in the invisible resources/visible foragers condition, there was limited evidence that foragers paid attention to others' locations in order to find the pools at the beginning of the experiment, and then they ignored other foragers for the rest of the experiment. This strategy may be analogous to findings that guppies rely on personal experience when they are familiar with the environment, but the rely on others as information cues when the environment is unfamiliar (Kendal, Coolen, and Laland, 2004). Furthermore, evidence in other species, such as starlings, indicates that members gauge the success of others in order to decide where to forage (Templeton and Giraldeau, 1996). Galef Jr. and Giraldeau (2001) posit that many vertebrate species forage using a producer scrounger system. In this system, a producer discovers a food source and exploits it. The scrounger realizes that the producer has a successful patch of resources and therefore exploits the same patch.

The current experiment examines the extent to which human foragers use each other as sources of information. The experiment extends the task used by Goldstone and Ashpole (2004) and Goldstone, et al. (2005) by using colors to dynamically indicate the success of other foragers. In addition to the indications of success, our experiment also shifts the environment in the middle of the experiment. Whereas the previous experiments tested groups in each condition for five minutes and maintained the pool distributions (e.g. an 80% pool and a 20% pool) throughout that time, our experiment shifts the pool distributions in each condition. Therefore, the pools start with 70% and 30% distributions, respectively, but these later shift to 30% and 70%. The shifting distribution allows us to examine how quickly the groups adapt to changes in the environment, and we can test whether the speed of adaptation varies according to the availability of public and private information. We hypothesize that foragers will use each other as sources of information when the environmental uncertainty increases. In particular, when the food is invisible and other foragers are visible, we predict that foragers will rely on each other's locations as information as they find less food due to the environmental shift. Moreover, we hypothesize that indications of foragers' success will influence subsequent matching behavior and lead to slower group adjustments to the changing environment because indications of success lag behind the environmental changes. Thus, foragers will see that some individuals still appear to be successful at the previously abundant pool despite apparent changes in the food availability.

## **METHODS**

Twelve groups of undergraduates at DePauw Unversity received course credit for participating. Group sizes ranged from 7 to 12 participants, with an average of 9.33 participants per group. Participants sat at respective computers in a university computer lab. The experiment was designed using the NetLogo programming language and the accompanying Hubnet Client to allow multiple people to simultaneously forage in a virtual world. Each participant was randomly assigned to a default Netlogo icon (e.g. heart, butterfly, dog, etc.), and participants moved their icons through a 60 X 60 gridworld by pressing the 'i', 'j', 'k', and 'l' keys to move up, left, down, and right. Participants' locations and number of food pellets eaten were recorded every two seconds.

Each group completed six games, and each game lasted five minutes. In all games, two resource pools were randomly placed at least 40 units apart in the gridworld. At each time step (1/10 of a second), there was a (5.25% * number of foragers) probability of placing a new piece of food in one pool and a (2.25% * number of foragers) probability of placing a new piece of food in the other pool. Thus, on average, 70% of the resources were entering one pool while 30% were entering the other pool. Food placement was uniformly distributed within eight spaces of the pool center. Because we wanted to examine how quickly groups would adapt to changes in the environment, each game included a time point at which the resources switched distributions, i.e. the respective pools switched their probabilities so that 70% of the new food entered the previous 30% pool and vice versa (old food pieces remained until foragers picked them up). We did not want participants to expect a pool switch to occur at a certain time point, so six possible switch points were spaced at 12 second intervals between 162 seconds and 210 seconds, thereby allowing sufficient time for us to assess the matching equilibrium before and after the switch. The pairing of switch times with experimental conditions was counterbalanced across participant groups; however, due to experimenter error, one pairing occurred three times and one pairing only occurred once.

The six games tested six out of eight permutations of our three independent variables: visibility of resources, visibility of foragers, and indication of success. When the resources were visible, each piece of food appeared as a green square on everyone's screens. When the resources were invisible, a piece of food would only appear on a participant's screen for two seconds if he or she had just stepped on the food and collected it. Thus, participants could gradually understand where the food was located in the world. When the foragers were visible, everyone could see each other's icons move around in the world. However, when the foragers were invisible, each person only saw his or her own icon on the screen. Finally, each forager's success could be indicated by coloring the participant's icon according to how much food he or she had collected in the previous 30 seconds: 0-5 pieces = blue; 6-10 pieces = yellow, 11-15 pieces = orange, more than 15 pieces = red. Importantly, these indications of success were updated in 30 second blocks rather than continuously. When success was indicated, each forager could therefore interpret the success of the other foragers by looking at the color of their icons. When success was not indicated, all participants' icons were purple. We excluded two conditions because it was unreasonable to pair invisible foragers with success indication. Condition order was randomized across groups

## **RESULTS**

Figure 1 shows the non-normalized matching results for the six conditions. The vertical line in each graph denotes the relative switch point, with all games aligned in order to compare matching behavior before and after the switch. Due to a network timing issue in NetLogo, one condition with a switch point at 210 seconds ended at 288 seconds, so each graph represents the window of time consisting of 78 seconds before the switch and 78 seconds after the switch.

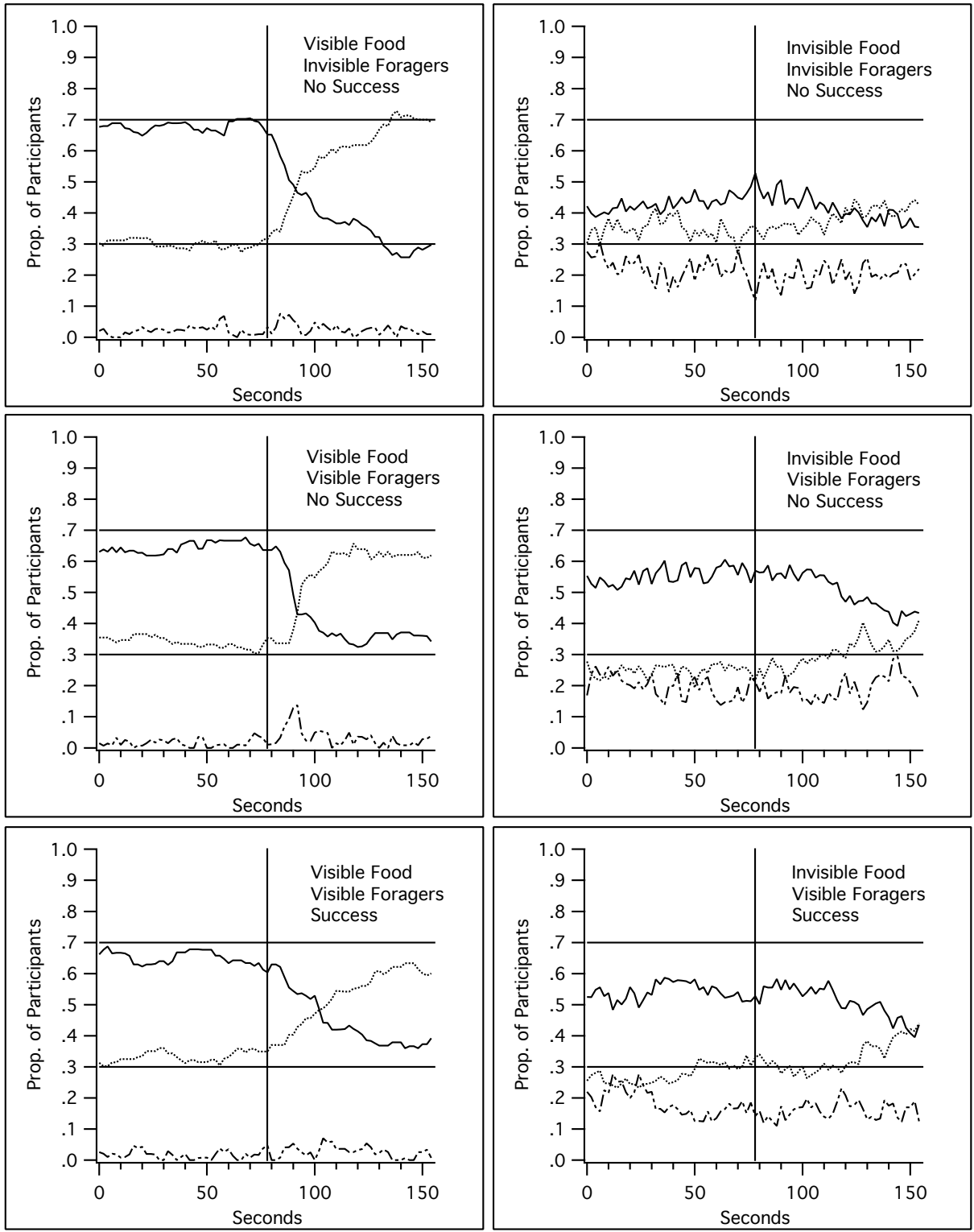

*Figure 1: Non-normalized matching results for the six conditions*

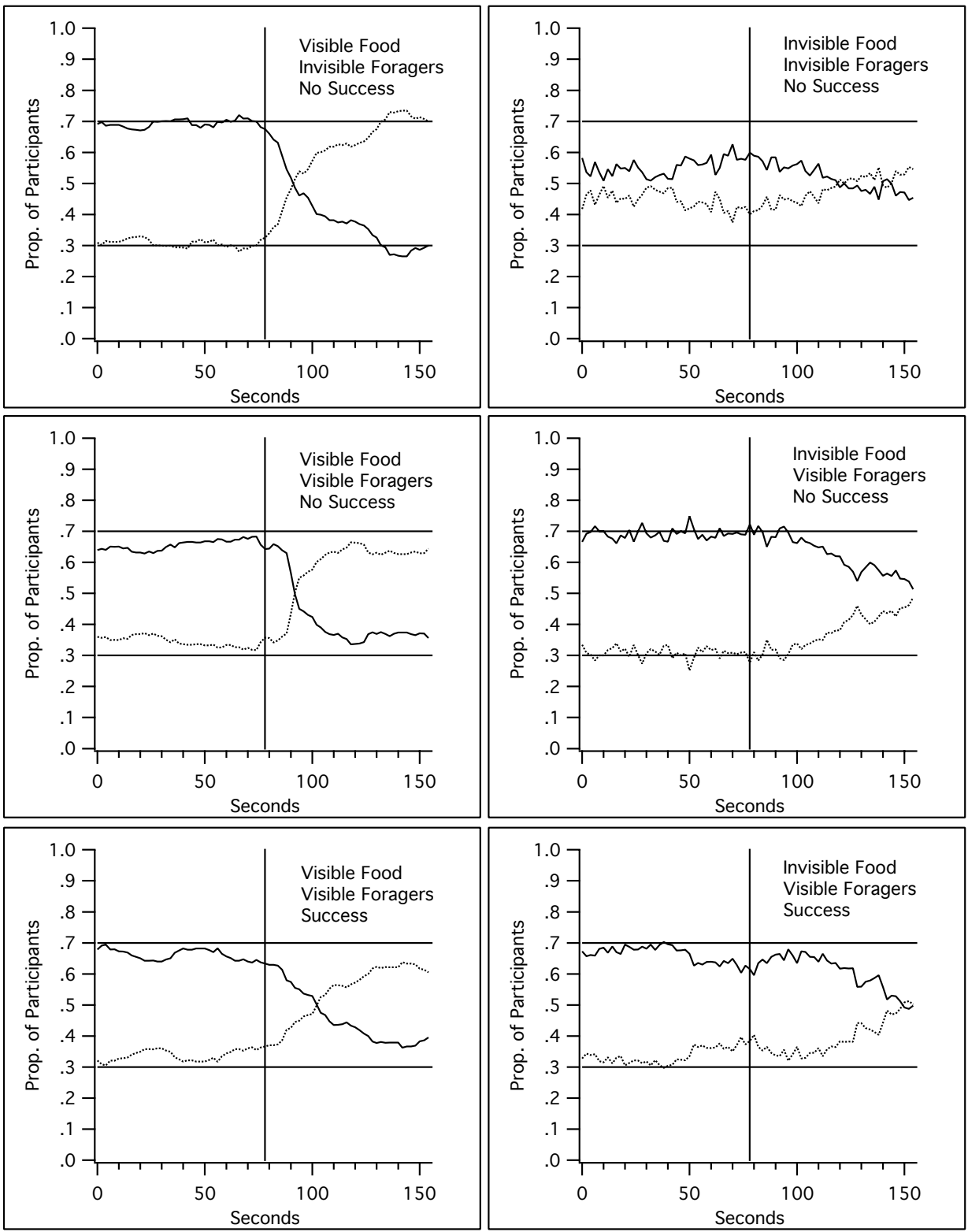

Figure 2: Normalized matching results for the six conditions

At each time step, a participant was classified as within a pool if he or she was within 13 units of the pool center (the actual pool periphery was 8 units, and a distance of 20 units would lead to an overlap for the minimum pool distance of 40). Each graph averages across the 12 groups to show the proportion of foragers in each pool across time. The bottom dashed line indicates the proportion of foragers who were outside of both pools, which suggests that the foragers were either switching pools or – when the food is invisible – searching for the pools. Figure 2 shows the normalized matching results, i.e.

$$\frac{\text{Proportion in Pool 1}}{(\text{Proportion in Pool 1 + Proportion in Pool 2})}$$

Notably, the results prior to the switch point in the top four graphs of Figure 2 largely replicate those reported in Goldstone and Ashpole (2004) and Goldstone, Ashpole, and Roberts (2005). The latter article found slight overmatching in the visible food/invisible foragers condition, whereas our results show nearly perfect matching in that condition. The visible food/visible foragers condition and the invisible food/invisible foragers condition both led to undermatching, with significantly greater undermatching in the latter condition as reported by Goldstone and Ashpole. Table 1 shows the average proportion of participants in the 70% pool for each condition 78 seconds before the switch. A 2 x 2 repeated measures ANOVA shows a marginal interaction, $F(1,11) = 4.28$, $p = .063$, and no main effects. Table 2 shows a similar comparison for the bottom four conditions, and a 2 (invisible vs. visible food) X 2 (no success vs. success) repeated measures ANOVA showed no main effects or interaction.

Given the inherent differences in undermatching between the six conditions, we compared each condition's matching behavior after the switch to its behavior before the switch. Thus, for each group in each condition, we averaged the proportion of participants in the (initially) 70% for the 78 seconds before the switch, and we subtracted the averaged proportion of participants in the pool for the 78 seconds after the switch. Table 3 shows the resulting difference values for the four conditions without success. A 2 (invisible vs. visible food) X 2 (invisible vs. visible foragers) repeated measures ANOVA showed a main effect for food visibility, $F(1,11) = 38.45$, $p < .001$, and a marginal interaction, $F(1,11) = 3.50$, $p = .088$. The main effect indicates that foragers adjust to the environmental change soon after the pool switch when the food is visible, but they fail to recognize and adjust to the environmental

|  | Invisible food | Visible food |
|---|---|---|
| Invisible foragers | .553 | .690 |
| Visible foragers | .675 | .640 |

*Table 1: Normalized proportion of foragers in 70% pool at 78 seconds before the environmental shift*

|  | Invisible food | Visible food |
|---|---|---|
| No success | .675 | .640 |
| Success | .672 | .678 |

*Table 2: Normalized proportion of foragers in 70% pool at 78 seconds before the environmental shift (all four conditions involve visible foragers)*

|  | Invisible food | Visible food |
|---|---|---|
| Invisible foragers | .031 | .302 |
| Visible foragers | .071 | .239 |

*Table 3: Differences between the normalized proportions of foragers in 70% pool before and after the environmental shift*

|  | Invisible food | Visible food |
|---|---|---|
| No success | .071 | .239 |
| Success | .054 | .205 |

*Table 4: Differences between the normalized proportions of foragers in 70% pool before and after the environmental shift (all four conditions involve visible foragers)*

changes when the food is invisible. However, the marginal interaction is driven by the reaction to visible foragers. When food is visible, the participants seem to be slightly deterred by the switching behavior of other foragers, but when the food is invisible, the greater switching in the visible foragers condition suggests that participants are using each other as sources of information about the changing environment.

Table 4 shows a similar analysis for the respective "no success" and success conditions. A 2 (invisible vs. visible food) X 2 (no success vs. success) repeated measures ANOVA only showed a main effect for food visibility, $F(1,11) = 77.37$, $p < .001$. Thus, indicating the recent success of foragers did not affect the speed with which groups adjusted to the changing environment.

## DISCUSSION

The results supported our first hypothesis that foragers would rely upon public information when the environment shifts. In the invisible food/visible agents condition, foragers trended towards switching pools more quickly than the invisible food/invisible foragers condition after the distributions changed. This suggests that foragers detect their changing personal payoffs and begin using others as cues to the environmental shift. However, when the food is visible, there is no indication that foragers used visible competitors as information sources when the environment changed. The null result suggests that foragers in these conditions relied on their own personal, perceptual assessments of the food throughout the game rather than using others as additional public information.

Our second hypothesis was not supported. We predicted that indications of success would actually decrease how quickly a group adjusts to the environment because indications of success lag behind the environmental changes. In these experiments, success was only updated every 30 seconds, so if foragers use each other's success as information, then the 30 second lag could have significantly delayed switching behavior after the environment changed. Instead, when the food was visible, the results provided further evidence that foragers only focus on the food rather than using other foragers as information. Perhaps the surprising null result occurs in examining the role of indications of success on the invisible food/visible foragers condition. As discussed above, foragers seem to use each other as information sources in this condition, but the indications of success do not seem to enhance this reliance on others. Foragers may discount others' short-term success in this condition because they realize that the environment is changing (hence the only reason to pay attention to others as information) and recognize that the details of others' current success may not be representative.

We are currently performing more detailed analyses regarding the comparative inequality of the conditions (as measured by Gini coefficients) and the efficiency of food collection in the conditions. It is possible that these analyses will reveal some distinctions in the use of public and private information that are not evident in the current focus on group matching behavior.

We are also working on an extension to the EPICURE model (Roberts and Goldstone, 2006) that can fit the current data. The original model lacks a parameter to weight the success of other foragers, but perhaps just as importantly, the model also does not incorporate reward history into foragers' behaviors when the food is visible. A reward history may be necessary in order to model the current results in which even the visible food conditions show a lag before the foragers switch pools in response to the changing environment. However, it is also possible that reward histories (and their accompanying bias to stay where you have recently been rewarded) will be unnecessary, and the lag may simply occur because it takes a while for the changing food distributions to be perceptually detectable and then promote switching behavior.

## REFERENCES

Fretwell, S. D., & Lucas, H. J. (1970). Ideal free distribution. *Acta Biotheory*, *19*, 16–21.

Galef Jr., B., & Giraldeau, L.A. (2001). Social influences on foraging in vertebrates: causal mechanisms and adaptive functions. *Animal Behavior*, *61*, 3-15.

Goldstone, R.L., & Ashpole, B.C. (2004). Human foraging behavior in a virtual environment. *Psychonomic Bulletin & Review*, *11*, 508-514.

Goldstone, R.L., Ashpole, B.C., & Roberts, M.E. (2005). Knowledge of resources and competitors in human foraging. *Psychonomic Bulletin & Review, 12*, 81-87.

Kendal, R. L., Coolen, I., & Laland, K. N. (2004). The role of conformity in foraging when personal and social information conflict. *Behavioral Ecology, 15*, 269–277.

Roberts, M. E., & Goldstone, R. L. (2006). EPICURE: Spatial and Knowledge Limitations in Group Foraging. *Adaptive Behavior, 14*, 291-313.

Templeton, J. J., & Giraldeau, L. A. (1996). Vicarious sampling: The use of personal and public information by starlings foraging in a simple patchy environment. *Behavioral Ecology and Sociobiology, 38*, 105–114.